\documentclass[12pt,a4paper]{article}
\usepackage{fullpage}
\usepackage[centertags]{amsmath}
\allowdisplaybreaks[4]
\usepackage{amssymb}
\usepackage{bm}
\usepackage{url}
\usepackage[dvips,hyperindex]{hyperref}
\begin{document}

\begin{flushright}
\setlength{\tabcolsep}{2pt}
\begin{tabular}{rl}
Main Text: & 30 Jan 2008
\\
Addendum 1: & 4 Mar 2008
\\
Addendum 2: & 17 Apr 2008 (revised)
\end{tabular}
\end{flushright}
\vspace{1cm}
\begin{center}
\Large\bfseries
Comment on the Neutrino-Mixing Interpretation of the GSI Time Anomaly
\\[0.5cm]
\large\normalfont
Carlo Giunti
\\[0.5cm]
\normalsize\itshape
\setlength{\tabcolsep}{1pt}
\begin{tabular}{cl}
INFN, Sezione di Torino,
Via P. Giuria 1, I--10125 Torino, Italy
\end{tabular}
\end{center}
\begin{abstract}
It is shown that 
neutrino mixing cannot explain the GSI time anomaly,
refuting recent claims in this direction.
Addendum 1: Remarks on \texttt{arXiv:0801.1465}.
Addendum 2: Quantum effects in GSI nuclear decay.
\end{abstract}

A GSI experiment observed an oscillatory time modulation
of the electron-capture decay of
$^{140}\text{Pr}^{58+}$
and
$^{142}\text{Pm}^{60+}$
ions,
\begin{equation}
{}^{140}\text{Pr}^{58+} \to {}^{140}\text{Ce}^{58+} + \nu_{e}
\,,
\qquad
{}^{142}\text{Pm}^{60+} \to ^{142}\text{Nd}^{60+} + \nu_{e}
\,,
\label{000}
\end{equation}
with periods
$T(^{140}\text{Pr}^{58+}) = 7.06(8) \, \text{s}$
and
$T(^{142}\text{Pm}^{60+}) = 7.10(22) \, \text{s}$
\cite{0801.2079}.
The experimental collaboration wrote:
``tentatively this observation is attributed to the coherent superposition
of finite mass eigenstates of the electron neutrinos from the weak decay into a two-body final state''
\cite{0801.2079}.
Theoretical arguments towards this interpretation have been presented in Refs.~\cite{0801.2121,0801.3262}.
Unfortunately,
this interpretation is in contradiction with the
well known fact that decay rates and cross sections
are given by the incoherent sum over the different channels
corresponding to different massive neutrinos
\cite{Shrock:1980vy,McKellar:1980cn,Kobzarev:1980nk,Shrock:1981ct,Shrock:1981wq,hep-ph/0608070,Giunti-Kim-2007}.
In this comment I would like to explain the mistake in the calculations presented in
Refs.~\cite{0801.2121,0801.3262}.

The authors of Refs.~\cite{0801.2121,0801.3262}
calculated the electron capture process using time-dependent perturbation theory
with the effective time-dependent weak interaction Hamiltonian
\begin{equation}
H_{\text{W}}(t)
=
\frac{G_{\text{F}}}{\sqrt{2}} \, V_{ud}
\int \text{d}^3x \,
\overline{\psi_{n}}(x) \gamma^{\mu} \left( 1 - g_{A} \gamma^{5} \right) \psi_{p}(x)
\sum_{j=1}^{3}
U_{ej}^{*}
\overline{\psi_{\nu_{j}}}(x) \gamma_{\mu} \left( 1 - \gamma^{5} \right) \psi_{e}(x)
\,,
\label{001}
\end{equation}
using standard notations.
They interpreted
(see Eq.~(3) in Ref.~\cite{0801.2121}
and
Eq.~(23) in Ref.~\cite{0801.3262})
\begin{equation}
A(t)
=
\sum_{k} A_{k}(t)
\label{002}
\end{equation}
as the time-dependent amplitude of the decay
\begin{equation}
I_{i} \to I_{f} + \nu_{e}
\,,
\label{003}
\end{equation}
where
\begin{equation}
A_{k}(t)
=
\int_{0}^{t} \text{d}\tau
\langle I_{f}, \nu_{k} | H_{\text{W}}(\tau) | I_{i} \rangle
\label{004}
\end{equation}
is the time-dependent amplitude of
\begin{equation}
I_{i} \to I_{f} + \nu_{k}
\label{005}
\end{equation}
transitions.
Here $I_{i}$ is the initial ion
($^{140}\text{Pr}^{58+}$
or
$^{142}\text{Pm}^{60+}$),
$I_{f}$ is the final ion
($^{140}\text{Ce}^{58+}$
or
$^{142}\text{Nd}^{60+}$,
respectively),
and
$\nu_{k}$
are the massive neutrinos ($k=1,2,3$).

Regrettably,
the amplitude in Eq.(\ref{002}) does not describe the decay (\ref{003}),
but a decay in which the final neutrino state is
\begin{equation}
| \nu \rangle
=
\sum_{k} | \nu_{k} \rangle
\,,
\label{007}
\end{equation}
which is clearly different from an electron neutrino state.
Indeed,
in the standard theory of neutrino oscillations
(see
Refs.~\cite{Bilenky:1978nj,Bilenky:1987ty,hep-ph/9812360,hep-ph/0202058,hep-ph/0306239,hep-ph/0608070,Giunti-Kim-2007})
electron neutrinos are described by the state
\begin{equation}
| \nu_{e} \rangle
=
\sum_{k} U_{ek}^* \, | \nu_{k} \rangle
\,,
\label{1331}
\end{equation}
where $U$ is the unitary mixing matrix of the neutrino fields in Eq.~(\ref{001}).
More accurately,
if the neutrino mass effects in the interaction processes are taken into account
\cite{Giunti:1992cb,hep-ph/0102320,hep-ph/0306239,hep-ph/0402217,hep-ph/0608070,Giunti-Kim-2007},
in the time-dependent perturbation theory used in Refs.~\cite{0801.2121,0801.3262}
the final electron neutrino in the process (\ref{003}) is described by the normalized state
\begin{equation}
| \nu_{e}(t) \rangle
=
\left( \sum_{j} |A_{j}(t)|^{2} \right)^{-1/2}
\sum_{k} A_{k}(t) \, | \nu_{k} \rangle
\,.
\label{1332}
\end{equation}
The time dependence of this electron neutrino state
takes into account the fact that in time-dependent perturbation theory
the final state of a process is studied during formation.

Using the correct electron neutrino state in Eq.~(\ref{1332}),
the decay amplitude is not given by
Eq.(\ref{002}), but by
\begin{equation}
A(t)
=
\left( \sum_{j} |A_{j}(t)|^{2} \right)^{-1/2}
\int_{0}^{t} \text{d}\tau
\sum_{k} A_{k}^{*}(t) \langle I_{f}, \nu_{k} | H_{\text{W}}(\tau) | I_{i} \rangle
=
\left( \sum_{k} |A_{k}(t)|^{2} \right)^{1/2}
\,.
\label{010}
\end{equation}
Then,
it is clear that the electron capture
probability is given by the incoherent sum over the different channels
of massive neutrino emission.
In other words,
there is no interference term between different massive neutrinos
contributing to
the rates of the electron capture processes in Eq.~(\ref{000}),
as well as all decay rates and cross sections.

In conclusion,
I have shown that neutrino mixing cannot explain the GSI time anomaly \cite{0801.2079},
refuting the claims presented in Refs.~\cite{0801.2121,0801.3262}.

\section*{Addendum 1: Remarks on \texttt{arXiv:0801.1465}}

This addendum is motivated by associations of Ref.~\cite{0801.1465}
with the GSI anomaly and the first version of this comment
(see for example the January 2008 issue of
\emph{Long-Baseline Neutrino Oscillation Newsletters}
at
\url{www.hep.anl.gov/ndk/longbnews/0801.html}).

The author of Ref.~\cite{0801.1465} predicted\footnote{
Ref.~\cite{0801.1465} is dated 9 January 2008,
but the work started in 2006 (see the Acknowledgments).
}
the possibility to observe oscillating decay probabilities
due to neutrino mixing.

Frankly speaking, the arguments presented in
Ref.~\cite{0801.1465}
seem to me rather obscure.
Furthermore,
no real calculation of the decay probability is presented.
Therefore,
I wish only to comment on the points
in Ref.~\cite{0801.1465}
which are relevant for the problem under discussion:

\begin{enumerate}

\item
At the beginning of Section~II.B of Ref.~\cite{0801.1465}:
``Both energy and momenta are conserved for each component of the wave 
packet which has a momentum $\vec P$ and energy $E$ in the initial state.''

There is no such conservation law.

In a scattering or decay process with particles described by plane waves,
the energies and momenta of all the particles have definite
values.
The total energy and momentum are exactly conserved.
Mathematically,
the conservation follows from the integration over space-time of a product of plane waves which
generates an energy-momentum delta function.
Obviously this is an approximation (very often very good).

In reality, since all processes are localized in space-time,
there is always an energy-momentum uncertainty.
In this case,
the interacting particles are described by wave packets.
Talking of exact energy-momentum conservation makes no sense,
since energy and momentum do not have definite values.
They are conserved only within their uncertainty.
Mathematically,
the energy-momentum delta function is replaced by
a factor which suppresses energy-momentum violations larger
than the energy-momentum uncertainty
(see Refs.\cite{hep-ph/9305276,hep-ph/0205014}).

Notice that in this case nothing can be said about
the behavior of each component of a wave packet.
A wave packet interacts as a whole.

\item
Before Eq.~(2.4) of Ref.~\cite{0801.1465}:
``The difference  in momentum $\delta p_\nu$ between the two
neutrino eigenstates with the same energy produces a small initial momentum
change $\delta P$ \ldots''.

This would be a violation of causality:
final states are determined by initial states,
not vice versa.

In any case,
what is the meaning of $\delta P$
if the initial state is described by a wave packet
which depends on the way in which the initial state has been prepared?

\item
The relevant discussion in Section~II.C of Ref.~\cite{0801.1465}
is based on $\delta P$, which is physically meaningless.

\end{enumerate}

\section*{Addendum 2: Quantum effects in GSI nuclear decay}

In the first version of this addendum I incorrectly claimed that
the GSI anomaly cannot be due to a quantum effect in nuclear decay.
I would like to thank Yu.A. Litvinov for an enlightening discussion on this point
at the IV International Workshop on: ``Neutrino Oscillations in Venice''
(15-18 April 2008, Venice, Italy).

Although the ions are monitored with a frequency of
the order of the revolution frequency in the ESR storage ring, about 2 MHz,
the GSI anomaly could be due to the quantum interference between two coherent states of the
decaying ion
if the interaction with the measuring apparatus does not distinguish
between the two states.
In order to produce quantum beats with the observed period of about $ 7 \, \text{s} $,
the energy splitting between the two states must be of the order of
$ 10^{-16} \, \text{eV} $.
It is very likely that the measuring apparatus which monitors the ions
in the ESR storage ring does not distinguish
between these two states and their coherence is preserved for a long time.

The problem is to find the origin of such a small energy splitting.
The authors of Ref.~\cite{0801.2079}
noted that the splitting of the two hyperfine $1s$ energy levels
of the electron is many order of magnitude larger
(and the contribution to the decay of one of the two states is
suppressed by angular momentum conservation).
It is difficult to find a mechanism which produces a smaller
energy splitting.

\section*{Acknowledgments}

I would like to thank the Department of Theoretical Physics of the University of Torino
for hospitality and support.

\raggedright

\end{document}